\documentclass{desyproc}

\begin{document}
\title{cuLGT: Lattice Gauge Fixing on GPUs}

\author{{\slshape Hannes Vogt$^1$, Mario Schr\"ock$^2$}\\[1ex]
$^1$Institut f\"{u}r Theoretische Physik, Universit\"at T\"ubingen, Germany\\
$^2$INFN Roma Tre, Italy }

\contribID{27}

\confID{7534}  
\desyproc{DESY-PROC-2014-05}
\acronym{GPUHEP2014} 
\doi  

\maketitle

\begin{abstract}
We adopt CUDA-capable Graphic Processing Units (GPUs) for Landau, Coulomb and maximally Abelian gauge fixing in 3+1 
dimensional SU(3) and SU(2) lattice gauge field theories. A combination of simulated annealing and overrelaxation is 
used to aim for the global maximum of the gauge functional. We use a fine grained degree of parallelism to achieve the 
maximum performance: instead of the common 1 thread per site strategy we use 4 or 8 threads per lattice site. Here, we 
report on an improved version of our publicly available code (www.cuLGT.com and github.com/culgt) which again increases 
performance and is 
much easier to include in existing code. On the GeForce~GTX~580 we achieve up to 470 GFlops (utilizing 80\% of the 
theoretical peak bandwidth) for the Landau overrelaxation code.
\end{abstract}

\section{Introduction}
In lattice QCD, gauge fixing is necessary to study the fundamental (gauge-variant) two point functions of QCD and 
to compare with continuum results. Lattice gauge fixing is an optimization problem with very many degrees of freedom 
and many local maxima. Each local maximum corresponds to a so-called Gribov copy. These Gribov copies may have an 
effect on gauge-variant quantities. One way to get a unique gauge copy is to search for the global maximum of the 
functional. This task is very time-consuming and an efficient implementation on modern hardware is desirable. 
Since the optimization algorithms are nicely parallelizable, graphic processing units (GPUs) are perfectly suited for 
these algorithms. Here, we report on latest improvements to cuLGT, a CUDA-based software 
for lattice gauge fixing that evolved from the first GPU gauge fixing code presented in~\cite{Schrock:2011hq}. In its 
first version, cuLGT offered standalone applications for Landau, Coulomb and maximally Abelian 
gauge fixing in 3+1 dimensional SU(3) gauge theories using a combination of overrelaxation and simulated 
annealing~\cite{Schrock:2012fj}. One of the aims of cuLGT2 was to offer the possibility to integrate the gauge fixing 
routines in 
existing lattice QCD frameworks, like the MILC code~\cite{MILCCode}. In the following, we will restrict the discussion 
to Landau gauge and the overrelaxation algorithm. For a more complete treatment we refer to the original 
work~\cite{Schrock:2012fj}.

An alternative GPU gauge fixing software using a Fourier accelerated steepest descent algorithm is available from the 
authors of~\cite{Cardoso:2012pv}.

\section{Lattice Gauge Fixing}
On the lattice, a link variable $U_\mu(x) \in \text{SU}(N_c)$ transforms with respect to gauge transformations $g(x) 
\in \text{SU}(N_c)$ as 
\begin{equation}
\nonumber
 U_\mu \rightarrow U_\mu^g = g(x) U_\mu(x)g^\dag(x+\hat{\mu}).
\end{equation}
The continuum Landau gauge condition,
\begin{equation}
\nonumber
 \partial_\mu A_\mu(x) = 0,
\end{equation}
translates on the lattice to the discrete derivative
\begin{equation}
\label{eq:lattice:landaugauge}
 \Delta(x) = \sum_\mu \left( A_\mu(x)-A_\mu(x-\hat{\mu}) \right) = 0,
\end{equation}
where the connection between the continuum gauge fields $A_\mu(x)$ and the lattice links is established by 
\begin{equation}
\nonumber
 A_\mu(x) = \frac{1}{2\mathrm{i} a g}\left[ U_\mu(x)-U_\mu^\dag(x) \right]_\text{traceless}.
\end{equation}
In each local maximum of the Landau gauge fixing functional
\begin{equation}
\label{eq:lattice:landaufunctional}
 F^U[g] = \frac{1}{N_d N_c V}\sum_{x}\sum_{\mu} \mathrm{Re}\, \mathrm{tr} \left[ g(x) U_\mu(x) 
g^\dagger(x+\hat{\mu})\right]
\end{equation}
the lattice Landau gauge condition \eqref{eq:lattice:landaugauge} is satisfied. In the normalization of 
\eqref{eq:lattice:landaufunctional}, $N_d = 4$ denotes the number of space-time dimensions and $V=N_s^3N_t$ is the 
lattice volume. Instead of 
considering the complete functional \eqref{eq:lattice:landaufunctional}, we rewrite it in a sum of local terms by 
factoring out the gauge transformation at lattice point $x$,
\begin{equation}
\nonumber
 F^U[g] = \frac{1}{2N_d N_c V}\sum_{x} \mathrm{Re}\,\mathrm{tr} \left[ g(x) K(x) \right].
\end{equation}
Then, we optimize
\begin{equation}
\label{eq:lattice:localfunctional}
\mathrm{Re}\,\mathrm{tr} \left[g({x})K({x})\right] = \mathrm{Re}\,\mathrm{tr}\left[g({x}) \sum_\mu 
\left[ U_\mu({x}) + U^\dagger_\mu({x}-\hat{\mu}) 
 \right] \right].
\end{equation}
with respect to $g({x})$. All other (inactive) gauge transformations are set to the identity.
The local functional \eqref{eq:lattice:localfunctional} only depends on links that start or end at lattice site 
${x}$.

The local maximum at each site can be found directly as
\begin{equation}
\nonumber
 g(x) = K^\dag(x)/ \sqrt{\det{K^\dag(x)}}
\end{equation}
for the gauge group SU(2). For $N_c > 2$ one iterates over the SU(2) subgroups. To overcome the problem of critical 
slowing down, the authors of~\cite{Mandula:1990vs} proposed to apply an 
overrelaxation update by replacing $g(x)$ by $g^\omega(x)$ with $\omega \in [1,2)$. Since only transformations at 
neighboring sites interfere, we can use a \emph{checkerboard} layout and sweep first over the \emph{black} and then 
over the \emph{white} lattice sites.
The algorithm is then iterated until the discrete gauge condition \eqref{eq:lattice:landaugauge} is satisfied up to a 
given numerical precision.

\section{GPU optimizations}
Most GPU applications in lattice QCD are bound by the bandwidth of global device memory. Therefore, the highest 
focus should be on ideal memory usage. For an overview of optimization techniques for lattice QCD we refer 
to~\cite{Clark:2009wm}. In the following, we will shortly summarize the optimizations that led to the performance of 
cuLGT1. In Sec.~\ref{sec:cuLGT2} we will report on the improved code cuLGT2.

\subsection{Pre-cuLGT (first version)}

For maximal throughput the memory access should be coalesced, i.e. consecutive threads should read from
consecutive memory addresses. Therefore, we first reorder the lattice into its black and white sublattices according 
to the checkerboard layout. Within each sublattice we order data such that the site index runs faster than Lorentz 
and matrix indices.

In order to save bandwidth we do not load full $N_c \times N_c$ matrices from global memory, but only parts of it. 
Then, we use the properties of the gauge group to reconstruct the full matrix in registers/local memory. For the gauge 
group SU(3) we use a 12 parameter representation, i.e.\ the first 2 rows of the matrix.

\begin{figure}[ht]
\centerline{\includegraphics[width=0.75\textwidth]{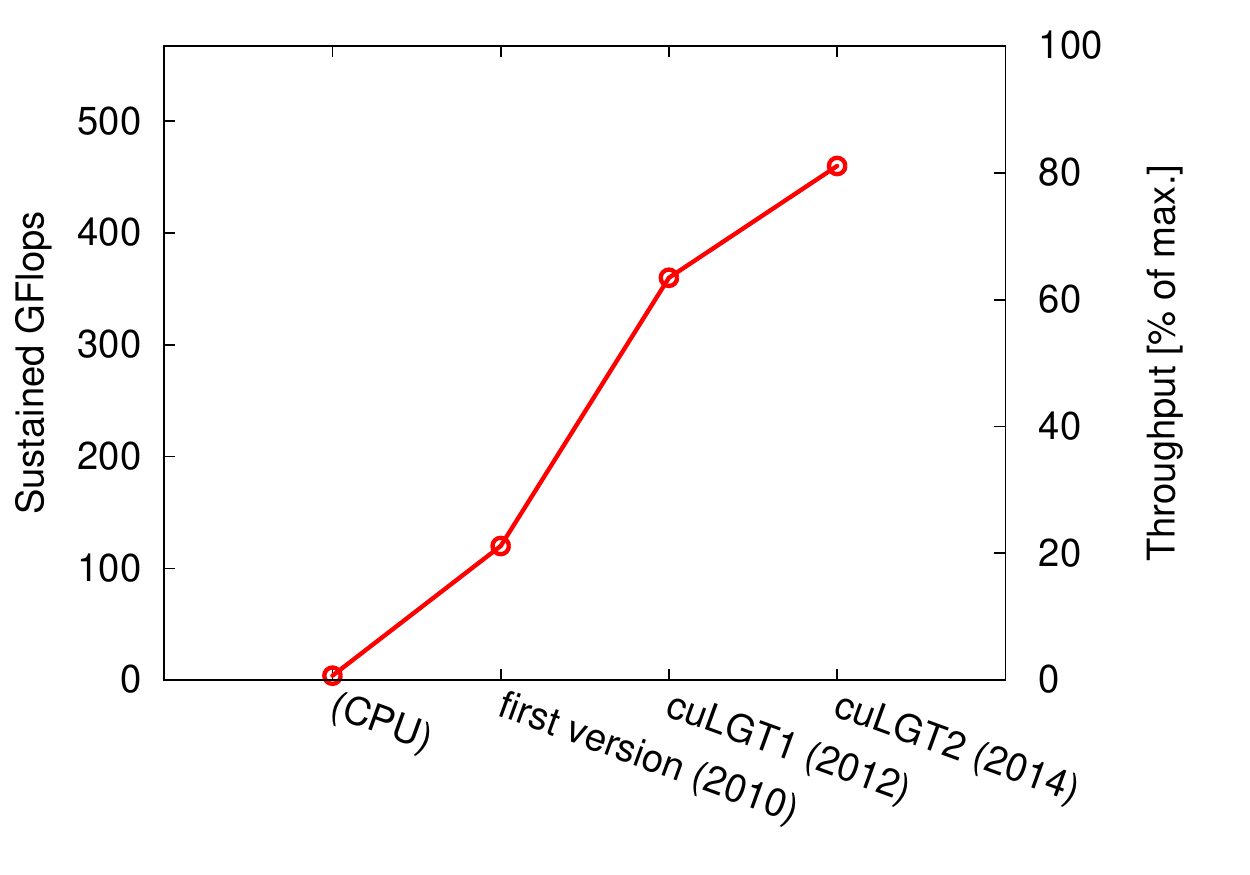}}
\caption{Evolution of the performance of cuLGT from the first version to cuLGT2 for the 
SU(3) Landau gauge overrelaxation code ($32^4$ lattice in single precision) on a 
GeForce~GTX~580. The CPU performance is from a single core of an Intel Core 
i7-2600.}\label{fig:performancehistory}
\end{figure}

With these optimizations we already get a remarkable speedup of a factor of $\sim30$ over a single core CPU 
implementation\footnote{We used our own reference CPU code which runs slightly faster than the 
publicly available MILC code~\cite{MILCCode}. However, for a \emph{fair} comparison of CPU~vs.~GPU performance we would 
need a highly optimized multi-threaded CPU code.} for the SU(3) overrelaxation code. The performance of 120 GFlops is 
of course far away from the theoretical peak performance of this GPU, however the correct measure of the utilization of 
the GPU is the achieved memory throughput. Therefore, on the r.h.s.\ of Fig.~\ref{fig:performancehistory} we show the 
throughput in percent of the theoretical peak bandwidth. For this version of the code we only use 20\% of the 
theoretical bandwidth\footnote{With the 12 parameter representation.}.

\subsection{cuLGT1}
Register pressure turned out to be the main performance bottleneck in the first version of the code. There we 
used one thread to calculate the update of a site. A simple calculation shows that we would already need 144 registers 
for storing the 8 links that are involved in a site update (and additional registers are needed for temporary storage in 
the computation). Since the register limit of GPUs of the Fermi generation is 63 registers (32 bit) per kernel, a 
lot of data is spilled to the slow global memory.
To relax the register pressure we introduced a 8-thread-per-site strategy in~\cite{Schrock:2012fj}, where we keep one 
link (18 parameters) in each thread. The communication among these 8 threads (summation of links to get $K(x)$ and 
distribution of the result $g(x)$) is done via shared memory. With this optimization we increase performance by a factor 
of three, using more than 60\% of the bandwidth. This version of the code is currently available on our website.

\section{cuLGT2}
\label{sec:cuLGT2}
With the development of cuLGT2 we wanted to solve several structural problems of cuLGT1: (a) the parameterization of 
links was hard-coded, switching to other parameterization would have needed a lot of code changes;  (b) related to the 
former point, SU(2) and SU(3) implementations needed a lot of code duplication; (c) switching from the 
8-threads-per-site strategy to 4 threads per site was not easily possible; (d) all these points prevented to implement 
a tool to automatically choose the optimal kernel settings for different GPUs, a technique that is already successfully 
used in the QUDA library~\cite{QUDALibrary}.

To systematically solve these issues we decided to completely rewrite major parts of the code. In the following, we 
leave out most details of the implementation but focus only on the link management. This part might be 
useful for many other lattice GPU applications, since it allows developers to use high level constructs for writing 
GPU optimized code.

\subsection{Link management}
Already in cuLGT1 we used two separate classes to represent a link in global memory (\texttt{SU3<Link>}) and local 
memory (\texttt{SU3<Matrix>}). The former takes care of the memory pattern (details about available memory 
patterns in~\cite{Schrock:2012fj}) and allows to access a link by specifying the site and the Lorentz index. The data 
is stored in a linear \texttt{float} or \texttt{double} array of length $L = VN_d(2N_c^2)$, where $2N_c^2$ is the 
number of parameters of the $N_c\times N_c$ complex matrix. For using the 12 parameter representation one just 
reads/stores the first two rows of the matrix. Other representations where the parameters are not just a subset of the 
full matrix, like the minimal 8 parameter representation, are not provided. Changing the datatype of the global array 
was also not possible\footnote{Using for example \texttt{float4}/\texttt{double2} instead of 
\texttt{float}/\texttt{double} would improve bandwidth utilization for memory accesses that cannot be coalesced, like 
access to links that are neighbors of the current site $U_\mu(x+\hat{\mu}$).}.

With cuLGT2, we decided to introduce an additional abstraction that easily allows changing how links are represented in 
memory. An implementation of a \texttt{ParameterizationType} defines the datatype and the number of elements. For 
transformations from one representation to another we define a mediator that overloads the assignment operator for the 
specific types. Links in global memory (\texttt{GlobalLink}) are now defined with two template parameters to specify 
the memory \emph{pattern} and the \emph{parameterization}. The \texttt{LocalLink} takes only the 
\emph{parameterization} type as template parameter. 
Two examples for \texttt{ParameterizationType}s for SU(3) are 
\vspace{5pt}

\hspace{-.5cm}
\begin{minipage}[t]{.49\textwidth}
\begin{itemize}
  \setlength{\itemsep}{-5pt}
 \item[] \texttt{class SU3Full}
 \item[] \verb|{|
 \item[] \hspace{.5cm}\texttt{typedef float TYPE;}
 \item[] \hspace{.5cm}\texttt{const static int SIZE = 18;}
  \item[] \verb|}|
\end{itemize}
\end{minipage}
\hfill
\noindent
\begin{minipage}[t]{.49\textwidth}
\begin{itemize}
  \setlength{\itemsep}{-5pt}
 \item[] \texttt{class SU3Vector4}
 \item[] \verb|{|
 \item[] \hspace{.5cm}\texttt{typedef float4 TYPE;}
 \item[] \hspace{.5cm}\texttt{const static int SIZE = 3;}
  \item[] \verb|}|
\end{itemize}
\end{minipage}

\vspace{5pt}
\noindent 
on the l.h.s.\ a 18 parameter representation with floats, 
usually used in \texttt{LocalLink}; on the r.h.s.\ a 12 parameter representation as three float4, usually used in 
\texttt{GlobalLink}.
A (simplified) example code to perform a gauge transformation is

\vspace{5pt}
\begin{minipage}{\textwidth}
\ttfamily
\begin{itemize}
  \setlength{\itemsep}{-5pt}
 \item[1] typedef GlobalLink<GPUPattern,SU3Vector4> GLOBALLINK;
 \item[2] typedef LocalLink<SU3Full> LOCALLINK;
 \item[3] void transformLink(Site s, int dir, LOCALLINK gLeft, LOCALLINK gRight)
 \item[4] \verb|{|
 \item[5] \hspace{.5cm}GLOBALLINK global(s, dir);
 \item[6] \hspace{.5cm}LOCALLINK local;
 \item[7] \hspace{.5cm}local = global;
 \item[8] \hspace{.5cm}local = gLeft*local*gRight.hermitian();
 \item[9] \hspace{.5cm}global = local;
 \item[10] \verb|}|
\end{itemize}
\end{minipage}

\vspace{5pt}
\noindent In line 1 and 2 the parameterizations for the \texttt{GlobalLink} and \texttt{LocalLink} are defined. 
Changing the gauge 
group to SU(2) would only require to set appropriate SU(2) parameterizations here.
In line 7 a \texttt{GlobalLink} is 
assigned to a \texttt{LocalLink}. The full matrix is implicitly reconstructed.
In line 8 the link is transformed. 
\texttt{LocalLink} overloads the multiplication operator and defines a function to compute the hermitian conjugate. The 
actual implementation of these operations is in the class of the \texttt{ParameterizationType}.
In line 9 the modified link is written back to global memory, discarding the third line.

\subsection{Performance}
\begin{figure}[ht]
\centerline{\includegraphics[width=0.75\textwidth]{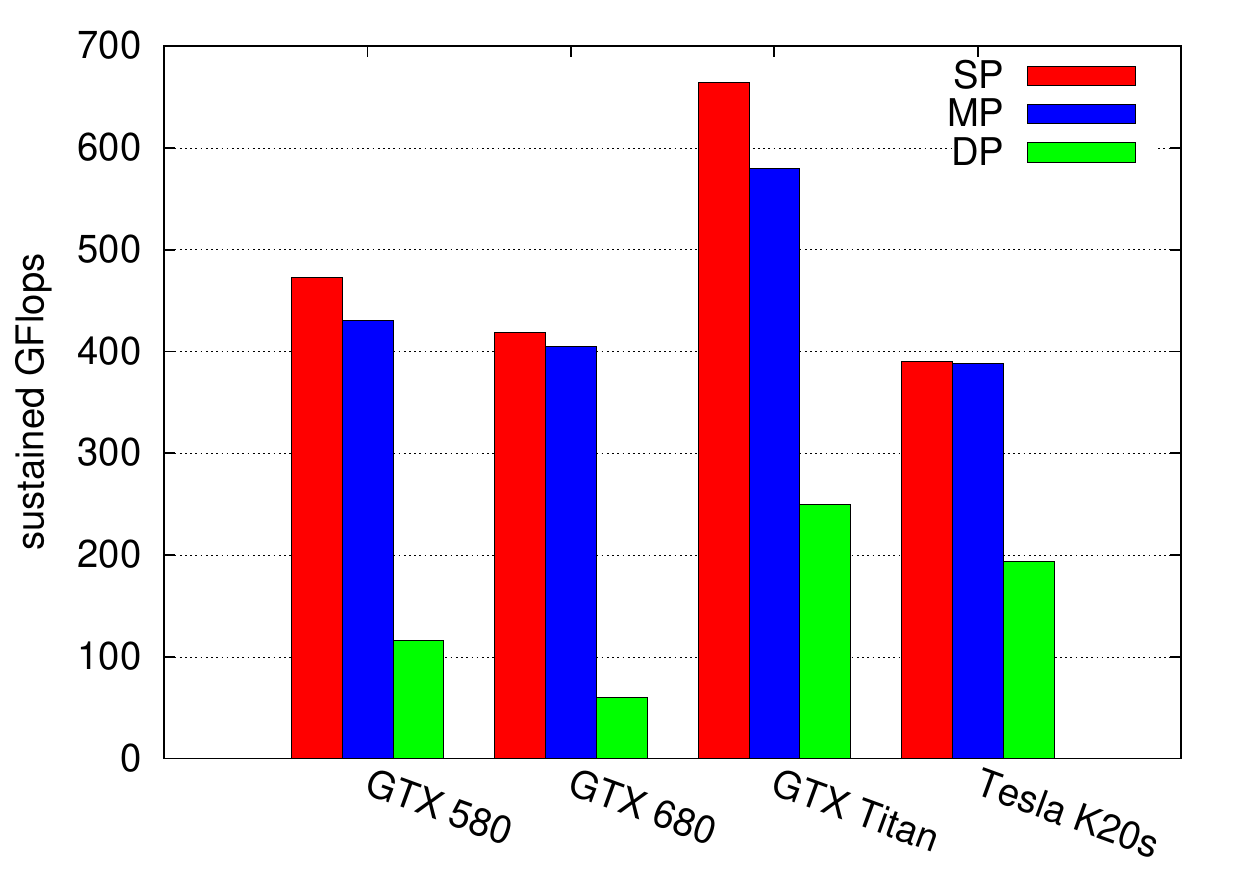}}
\caption{Performance of the cuLGT2 SU(3) Landau gauge overrelaxation code ($32^4$ lattice) on different 
GPUs in SP (left), mixed precision (middle) and DP (right)}\label{fig:gpucomparison}
\end{figure}

Although the primary design goal of cuLGT2 was not on performance improvements, we got a speedup compared 
to cuLGT1. The main 
improvement in Fig.~\ref{fig:performancehistory} comes from the use of the 4-threads-per-site strategy instead of 8 
threads per site. In Fig.~\ref{fig:gpucomparison} we compare the performance of the code on different GPUs. Only on the 
Tesla K20s 8-threads-per-site performs better. The moderate result indicates that we need additional tuning for the 
Tesla K20s.

\section{Summary}
With the development of cuLGT2 we successfully solved several design issues of cuLGT1. The gauge fixing software is now 
well modularized which allows us to run the gauge fixing routines from the MILC code. Additionally, the software 
automatically chooses the optimal kernel setup for different architectures at runtime by trying (a) 4 or 8 threads per 
site update, (b) different register limits by setting \texttt{\_\_launchbounds()} (c) switching texture loads on or 
off. With the 4-threads-per-site strategy and the improved link management the performance of the code was remarkably 
improved. The improved code will be available shortly on our website 
\url{cuLGT.com} and on \url{github.com/culgt}.

\section*{Acknowledgments}

H.V. acknowledges support from the Evangelisches Studienwerk Villigst e.V.

 

\begin{footnotesize}
\bibliographystyle{unsrt}
\bibliography{vogt_hannes.bib}


%

\end{footnotesize}


\end{document}